\documentclass[%
journal = nalefd,
manuscript = letter,
layout = twocolumn,
]{achemso}

\usepackage[utf8]{inputenc}

\usepackage{graphicx}
\usepackage{hyperref}
\usepackage{xr}

\usepackage{bm}
\usepackage{physics}
\usepackage{siunitx}
\usepackage{chemformula}

\usepackage{microtype}

\usepackage{xcolor}

\let\oldmaketitle\maketitle
\let\maketitle\relax

\newcommand{\kk}{\bm{k}}

\makeatletter
\newcommand*{\addFileDependency}[1]{
  \typeout{(#1)}
  \@addtofilelist{#1}
  \IfFileExists{#1}{}{\typeout{No file #1.}}
}
\makeatother

\newcommand*{\depexternaldocument}[2][]{%
    \externaldocument[#1]{#2}%
    \addFileDependency{#2.tex}%
    \addFileDependency{#2.aux}%
}

\title{Large Bulk Piezophotovoltaic Effect of Monolayer $2H$-\ch{MoS2}}
\date{\today}

\author{Aaron M. Schankler}%
\affiliation{Department of Chemistry, University of Pennsylvania, Philadelphia, Pennsylvania 19104--6323, USA}
\author{Lingyuan Gao}%
\affiliation{Department of Chemistry, University of Pennsylvania, Philadelphia, Pennsylvania 19104--6323, USA}
\author{Andrew M. Rappe}%
\affiliation{Department of Chemistry, University of Pennsylvania, Philadelphia, Pennsylvania 19104--6323, USA}
\email{rappe@sas.upenn.edu}

\depexternaldocument[SI-]{si}

\begin{document}

\twocolumn[{%
\begin{@twocolumnfalse}
\oldmaketitle
\begin{abstract}
The bulk photovoltaic effect in noncentrosymmetric materials is an intriguing physical phenomenon that holds potential for high-efficiency energy harvesting. Here, we study the shift current bulk photovoltaic effect in the transition-metal dichalcogenide \ch{MoS2}. We present a simple automated method to guide materials design and use it to uncover a distortion to monolayer $2H$-\ch{MoS2} that dramatically enhances the integrated shift current. Using this distortion, we show that overlap in the Brillouin zone of the distributions of the shift vector (a quantity measuring the net displacement in real space of coherent wave packets during excitation) and the transition intensity is crucial for increasing the shift current. The distortion pattern is related to the material polarization and can be realized through an applied electric field via the converse piezoelectric effect. This finding suggests an additional method to engineer the shift current response of materials to augment previously reported methods using mechanical strain.

\begin{description}
\item[Keywords]
BPVE, shift current, piezoelectric, first principles, monolayer, dichalcogenide
\end{description}
\end{abstract}
\end{@twocolumnfalse}
}]


In conventional photovoltaic devices, current is generated when the electric field created in a p-n junction separates electrons and holes. As they diffuse to the electrodes, these carriers lose energy and relax to the band edges through scattering events, which sets fundamental limits on the efficiency of junction-based photovoltaics \cite{Shockley1961}. In contrast, materials that lack inversion symmetry can generate photocurrent in the bulk \cite{Belinicher1980,Baltz1981,Belinicher1982}. This bulk photovoltaic effect (BPVE) was first observed in ferroelectric and piezoelectric materials \cite{Glass1974,Koch1976}, and because this process can generate above-band-gap photovoltages \cite{Glass1974}, devices using the BPVE are not bound by the same thermodynamic limits as junction-based devices \cite{Tan2016,Spanier2016}. The shift current is a quantum mechanical effect arising from off-diagonal elements of the density matrix \cite{Belinicher1988}. Recent experimental and theoretical studies indicate that shift current is a significant component of the BPVE \cite{Burger2019,Young2012,Fei2020}. Therefore, designing materials with enhanced shift current is a promising avenue for increasing the efficiency of photovoltaic devices. Several approaches have been studied experimentally to tune the shift current response of a material. The geometry of the sample and electrodes---for example using thin films or single-tip electrodes---plays a large role in the magnitude of the BPVE \cite{Spanier2016,Zenkevich2014}. Temperature \cite{Gong2018phonon}, strain \cite{Zhang2015piezopv,Nadupalli2019piezopv,Ai2020,Kaner2020}, and strain gradients \cite{Yang2018flexopv,Shu2020flexopv} can also significantly modulate the shift current. Applied strain can significantly improve photovoltaic device performance (termed the piezophotovoltaic effect). The photocurrent response of iron-doped \ch{LiNbO3} was increased by \SI{75}{\percent} under uniaxial compressive strain \cite{Nadupalli2019piezopv}, while strain increased both the open-circuit voltage and the short-circuit current of the multiferroic material \ch{BiFeO3}, enabling an increase of \SI{218}{\percent} in the power conversion efficiency \cite{Zhang2015piezopv}. Controlling photovoltaic response through applied strain gradients (analogously named the flexophotovoltaic effect) may produce a less dramatic response, but because the nonuniform strain breaks inversion symmetry, strain gradients can coerce the BPVE from otherwise inactive materials \cite{Yang2018flexopv}.

\begin{figure*}
\includegraphics{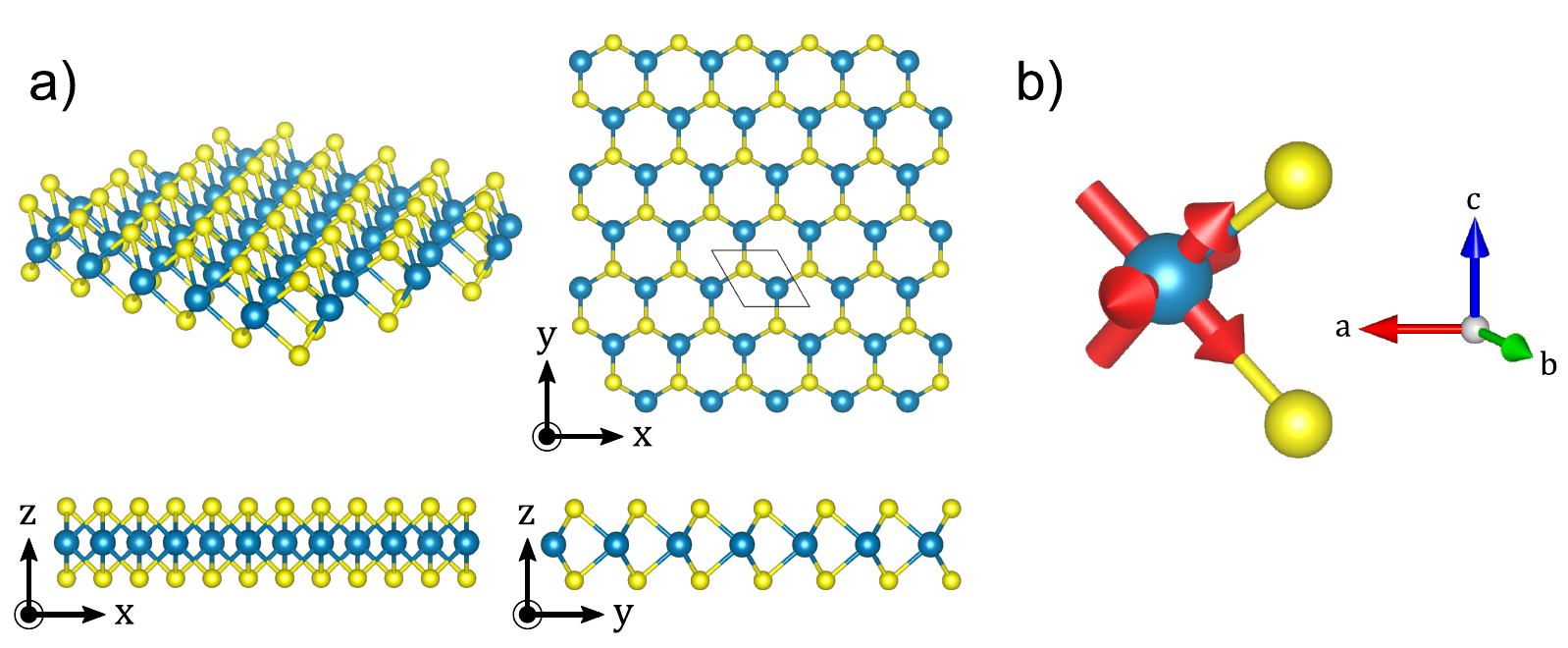}
\caption{\label{fig:structure} (a) $2H$-\ch{MoS2} crystal structure; (b) local Cartesian coordinates for \ch{Mo} displacement.}
\end{figure*}

Recent analytical work has connected the frequency integral of shift current and the polarization differences between filled and unfilled bands \cite{Fregoso2017scpol}. In monolayer \ch{GeSe}, the integrated shift current is linearly related to the total polarization at small displacements \cite{Rangel2017sctmmc}, while the relationship becomes more complex in materials with both positive and negative current responses at different frequencies such as \ch{PbTiO3} \cite{Young2012}, which makes total polarization unsuitable for material selection \cite{Tan2016}. In a study of photocurrent enhancement in \ch{Ni}-doped \ch{PbTiO3}, \citeauthor{Wang2016} showed that the shift current is mostly generated around a few points in $\kk$-space \cite{Wang2016}. They then connected the shift vector (a quantity related to the distance a carrier travels during excitation) at these points to the dopant distribution and artificial atom displacements. Eliminating shift vector cancellation between these points proved to be a viable method to enhance the shift current, suggesting that the shift vector could be a useful tool for materials design. Theoretical upper bounds have also shown that materials with a reduced band gap and more delocalized wave functions can have higher shift current \cite{Tan2019wfn,Tan2019bound}. A survey of many materials revealed that none approached this theoretical bound, suggesting an opportunity for materials design \cite{Tan2019bound}. Layered two-dimensional materials, and transition-metal dichalcogenides in particular, hold great promise in optoelectronic applications \cite{Bernardi2013,Koppens2014,Pospischil2014,Mak2016,Unuchek2018}. The transition-metal dichalcogenide \ch{MoS2} is a prototypical example of this class of materials with several phases exhibiting varied optical and electronic properties \cite{Zhao2018mos2phase,LopezSanchez2013,Fang2019}. The metastable $1T$ phase of \ch{MoS2} is inversion symmetric, but can spontaneously convert to various symmetry breaking ferroelectric phases \cite{Sharmila2014}. Shift current was predicted in the distorted $1T'''$ phase, where it was found to be tunable using biaxial strain \cite{Ai2020}. The $2H$ phase of monolayer \ch{MoS2} shown in Figure~\ref{fig:structure}a has a direct band gap, bright photoluminescence, and tunable electronic propenties \cite{Splendiani2010,Mak2016,Scalise2012,Conley2013,Lin2018}. It is noncentrosymmetric, so it can also exhibit nonlinear optical phenomena although the structure is nonpolar.

In this Letter, we outline an optimization method which can uncover atomic displacements that significantly enhance the single-frequency and integrated shift current responses in $2H$-\ch{MoS2}, and we show that overlap of the transition intensity and the shift vector in the Brillouin zone is crucial for large shift current. We connect a displacement identified through optimization with changes in the material polarization, and note that the distortion is realizable experimentally via the converse piezoelectric effect; thus, we name this phenomenon as ``the bulk piezophotovoltaic effect".

The shift current is a second-order optical effect and arises through a net displacement of excited electrons as the excited states evolve in an asymmetric crystal potential. There have been a variety of analytical descriptions of the shift current, including several that are amenable to numerical evaluation \cite{Young2012,Sipe2000,Wang2017wann,IbanezAzpiroz2018wann}. The frequency-dependent response tensor $\sigma$ may be split into two physically meaningful quantities:
\begin{align}\label{eq:shift}
    J_q &= \sigma_q^{rs} E_r E_s \nonumber\\
    \sigma_q^{rs} (\omega)
    &= e \sum_{m,n} \int \mathrm{d\kk}\, I^{rs}(m,n,\kk; \omega)\, R_q(m,n,\kk)
\end{align}
The quantity $I$ is the intensity of a transition and $R$ is the shift vector, which has units of length and describes the instantaneous travelling distance by carriers during excitation. These quantities are defined in (2) and (3), where $A$ is the Berry connection, $\phi$ gives the phase of the transition dipole, and $\omega_{mn}$ and $f_{mn}$ denote differences in band energies and fillings:
\begin{multline}
    I^{rs} (m,n,\kk; \omega) \\
    = \pi \pqty{\frac{e}{m\hbar\omega}}^2
    f_{nm} (\kk) \vb{P}_{mn}^r \vb{P}_{nm}^s \\
    \times\delta\bqty{\omega_{nm} (\kk) \pm\omega}, \\
\end{multline}
\begin{multline}
    R_q (m,n,\kk) \\ = \partial_{k_q} \phi_{mn} (\kk) - \bqty{A_{nq} (\kk) - A_{mq} (\kk)}.
\end{multline}
In this form, the shift current can be readily calculated from Bloch wave functions generated by standard density functional theory (DFT) calculation \cite{Young2012}. Calculation details are given in the Supporting Information.

We induce atomic displacements to probe how structural change influences the magnitude of shift current responses of monolayer $2H$-\ch{MoS2}. In one unit cell, there are nine degrees of freedom for ionic displacement. Previous study has revealed that the shift current is correlated with the wave function delocalization and the covalent bonding between neighboring atoms \cite{Tan2019wfn,Tan2019bound}. To capture this effect, we use a local Cartesian coordinate system aligned along the \ch{Mo-S} bond for atomic displacements. In this system, shown in Figure~\ref{fig:structure}(b) for \ch{Mo} displacement, the first orthogonal axis is aligned along the \ch{Mo-S} bond, while the other two are perpendicular to the bond with one axis contained in the $xy$-plane and the other with an out-of-plane component.

\begin{figure*}
    \includegraphics{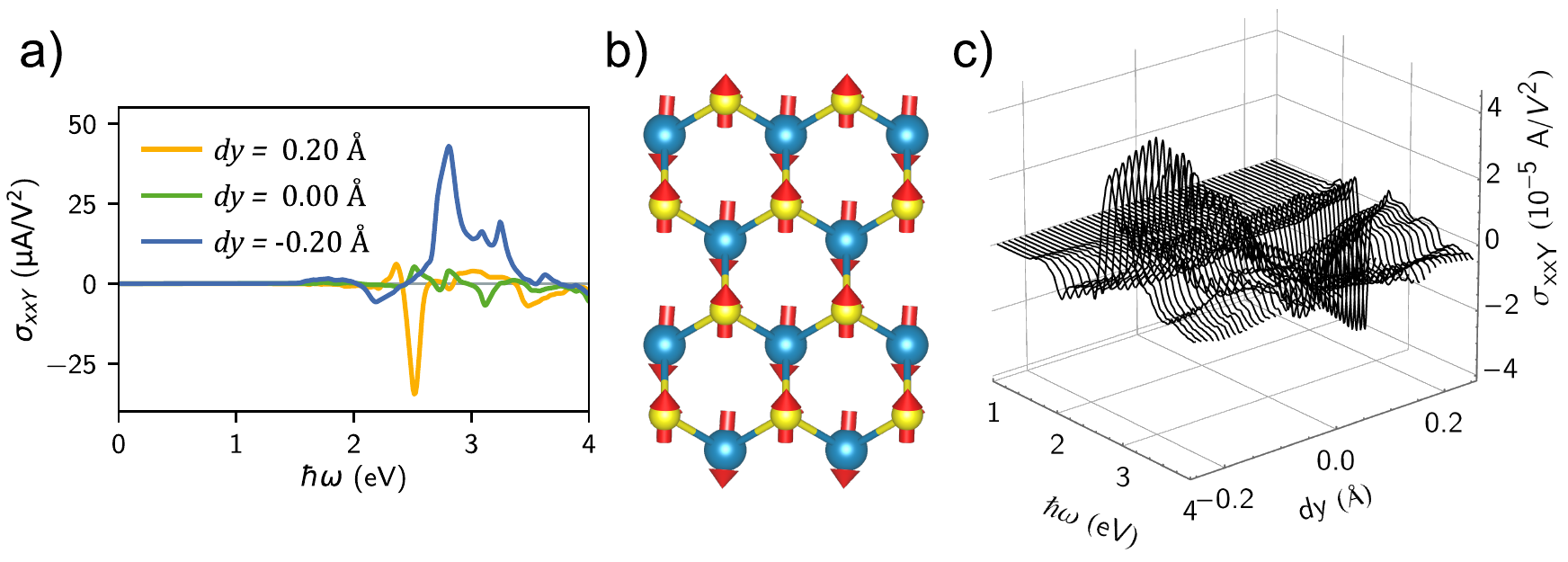}
    \caption{\label{fig:gradient} (a) $\sigma_{xxY}$ component of the \ch{MoS2} shift current tensor with \ch{Mo} displaced at $dy = \SI{0.2}{\angstrom}$ and $dy = \SI{-0.2}{\angstrom}$; (b) optimal displacement pattern; (c) smooth change in shift current response ($z$-axis) with continuous \ch{Mo} displacement in the $y$ direction.}
\end{figure*}

We use a gradient descent optimization scheme to reach the structure with the maximum shift current (SC) response. We focus on tensor elements with light polarized in the plane of the material, and focus specifically on the largest component, which is the $\sigma_{xxY}$ element. To optimize the structure, the shift current response is first calculated with small atomic displacements from a reference structure along the local Cartesian coordinates. Because SC is dependent on the electronic wave functions, which change continuously with ionic configuration, the SC response also varies smoothly with atomic displacement \cite{ibaezazpiroz2019quantitative} (for example see Figure~\ref{fig:gradient}c). This property allows the real space derivative of the frequency-dependent response tensor to be calculated using the finite-difference method in order to construct a linear approximation of the SC response as a function of atomic position. Using this approximation, the displacement of atoms with the steepest increase in the SC performance is identified. The SC performance is quantified using two different figures of merit: the maximum value of the response function over a specific frequency range and the integrated response over a frequency range. For photovoltaic applications, as the solar spectrum is continuous, the integrated response $\int \sigma^{rs}_{q}(\omega)d\omega$ is more relevant than a single-frequency response. The structure is optimized starting from the equilibrium structure by calculating the shift current gradient about a reference structure and then choosing a new reference based on this gradient. This procedure is applied iteratively to find high performing structures.

We present the optimal displacement pattern of $2H$-\ch{MoS2} in Figure~\ref{fig:gradient}b. The distortion is very simple; displacement is primarily confined to the $y$ direction, and when the \ch{Mo} atoms move forward towards one side of the triangle or backward towards a column of sulfur atoms on the vertex of the triangle, the maximal SC response can be realized. The photovoltaic tensor element $\sigma_{xxY}$ for the structure with \ch{Mo} displacements of \SI{0.2}{\angstrom} in the $\pm y$ direction are shown in Figure~\ref{fig:gradient}a. Compared to the reference structure, the maximum SC peak at \SI{2.8}{\eV} is increased five-fold and can reach \SI[per-mode = symbol]{40}{\uA\per\V\squared}. The integrated SC response in the frequency range of \SIrange[range-phrase = --, range-units = single]{0}{4}{\eV} is increased even more (more than ten-fold), because unlike in the equilibrium geometry, where the SC response has both positive and negative parts that will be largely cancelled after integration, responses in the optimal structure are mostly all the same sign, resulting in a larger integrated value. In contrast, biaxial strain of \SI{5}{\percent} triples the integrated response. Comparing the responses shown in Figure~\ref{fig:gradient}a, we note that the SC responses in the optimal structure are controllable. By reversing the displacement direction, the SC can also also be reversed, with a similar peak magnitude.

\begin{figure*}
    \includegraphics{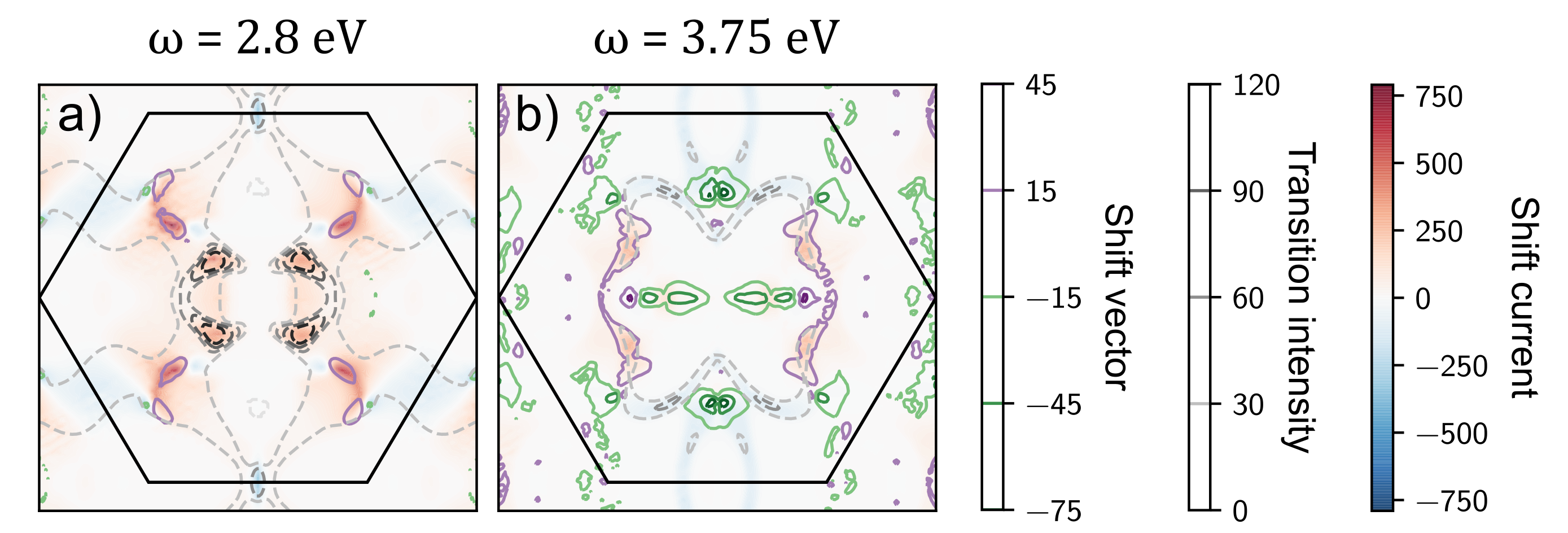}
    \caption{\label{fig:bz} $\kk$-resolved shift current, shift vector and intensity for transitions close in energy to  (a) $\omega = \SI{2.8}{\eV}$ and  (b) $\omega = \SI{3.75}{\eV}$. The shading indicates the $\kk$-resolved shift current, the dashed contours show the $\kk$-resolved intensity, and the solid contours show the $\kk$-resolved shift vector.}
\end{figure*}

To understand the large enhancement of SC response due to this simple displacement, we refer to Eq.~\ref{eq:shift}, which shows that the shift current is an integrated result over every $\kk$-point in the Brillouin zone (BZ). At each $\kk$-point, the contribution is a summation over band pairs of the product of the transition intensity and the shift vector (SV) of the corresponding band pair. For a 2D material, the contribution from the full BZ can be visualized. Following this idea, we plot the $\kk$-resolved transition intensity, shift vector, and SC separately, for transitions close to the $\omega = \SI{2.8}{\eV}$ peak (Figure~\ref{fig:bz}a). After displacing the Mo atom, the range of the transition intensity increases only slightly, but its distribution in the BZ is altered considerably (Figure~\ref{SI-fig:bz_tr}). In addition, more $\kk$-points with large shift vectors emerge after \ch{Mo} displacement (Figure~\ref{SI-fig:bz_sv}). With both increased intensity and shift vector, the range of SC variation is also increased. Inspecting Figure~\ref{fig:bz}a, the SC shares a similar distribution pattern with the shift vector. Peaks of $\kk$-resolved SC coincide with peaks in the $\kk$-resolved shift vector, but only in regions with appreciable transition intensity. This underscores a design principle of shift current that the enhancement requires a large overlap between the shift vector and transition intensity in the BZ. We also note that at these places, the shift vectors all have the same sign, so their contributions are not cancelled. As a contrast, the $\kk$-resolved signatures at $\omega = \SI{3.75}{\eV}$ are shown in Figure~\ref{fig:bz}b. Despite the large negative peaks in the shift vector, there are no corresponding peaks in the shift current because those peaks are not in regions of high transition intensity. Cancellation between positive and negative shift vectors at different locations in the BZ further reduces the net current after BZ integration. These two effects reduces the SC performance at this light frequency.

The sum of the transition intensity over occupied states is proportional to the imaginary part of the dielectric function $\epsilon^{bb}(\omega)$, and the BZ integral of the shift vector is related to the interband polarization difference $P_n - P_m$ with an additional gauge-dependent quantity related to $\phi_{mn}(\kk)$ \cite{Fregoso2017scpol}. Previous studies show that while geometric phase quantities like the shift current are only reproduced accurately with a full wave function description, the frequency-dependent dielectric function and the joint density of states are well approximated by a simple $k\cdot p$ model \cite{ibaezazpiroz2019quantitative}; therefore in Ref. \citenum{Fregoso2017scpol}, the dipole matrix elements were approximated as a constant in the BZ. This enabled the integrated SC response to be approximately written in the simplified form $\sum_{mn} \epsilon^{bb} (P_n - P_m)$, which is much simpler than equation \ref{eq:shift}. It directly relates the integrated SC response as a simple product of the frequency-dependent dielectric constant and the polarization difference of band pairs. However, as shown in Figure~\ref{fig:bz}, the approximation of a constant dipole matrix is not always well founded. Instead, the analysis in Figure~\ref{fig:bz} shows that the transition intensity can be highly nonuniform over the BZ.

\begin{figure}
\includegraphics{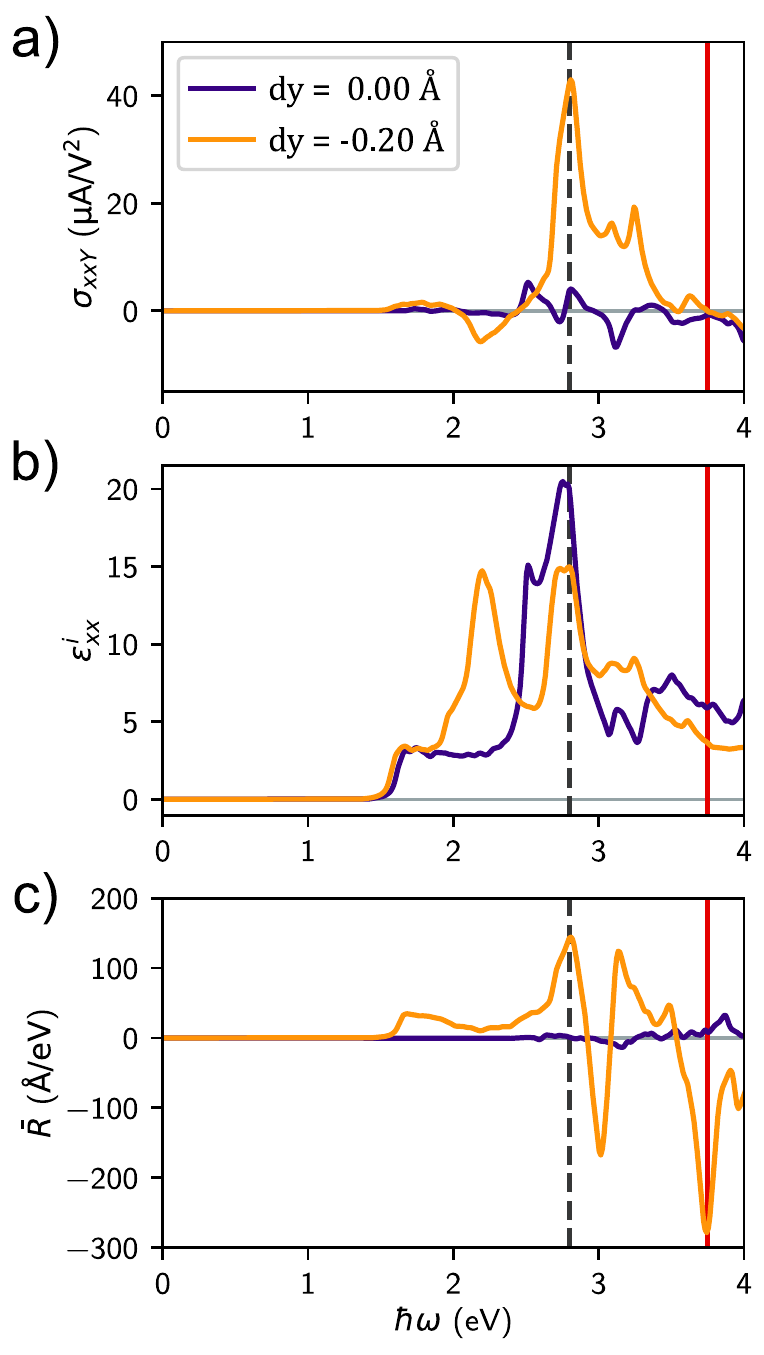}
\caption{\ch{MoS2} (a) photovoltaic tensor, (b) Brillouin-zone-averaged transition intensity, and (c) Brillouin-zone-averaged shift vector for both equilibrium and distorted structures.}
\label{fig:average}
\end{figure}

Although the dependence of the shift vector and the transition intensity on $\kk$ are crucial for fully explaining the shift current response, we can gain qualitative insight by examining these quantities averaged over the BZ at a single frequency. The aggregated shift vector $\bar{R}$, defined in Eq.~\ref{eq:avgSV}, is related to the polarization difference between conduction and valence bands, and the aggregated transition intensity is proportional to the imaginary dielectric function. 
\begin{equation}
\label{eq:avgSV}
    \bar{R}_q (\omega) = \sum_{m,n} \int \mathrm{d\kk} R_q (m,n,\kk) \delta(\omega_{mn} \pm \omega)
\end{equation}
These averaged quantities are plotted in Figure~\ref{fig:average} for both the reference structure and a distorted structure with \ch{Mo} displaced by $dy = \SI{-0.2}{\angstrom}$. The averaged transition intensity does not increase significantly in magnitude under distortion, while the averaged SV varies over a much wider range---at particular frequencies such as $\omega =$ \SIlist{2.8; 3.75}{\eV}, the averaged SV increases nearly 100-fold under atomic displacement. However, the increase in averaged shift vector does not guarantee the same increase in SC because the product of the averaged quantities is not equal to the SC response. For example, the response at $\omega = \SI{3.75}{\eV}$ is much weaker than that at $\omega = \SI{2.8}{\eV}$, although the averaged shift vector is larger for the former. This is consistent with the results shown in Figure~\ref{fig:bz}, where we showed that the misalignment between the shift vector and the transition intensity suppresses SC response associated with transitions that have large shift vectors. As a generalization of the SC design rule, we propose that the zone-averaged shift vector and intensity are good indicators of strong SC response. Since the shift vector can change more drastically, it is a more decisive factor. However, the overlap of transitions with large shift vector and high transition intensity at the same momentum is most important for determining the overall SC response, and this can be altered significantly with changed wave functions and different structures.

\begin{figure}
\centering
\includegraphics{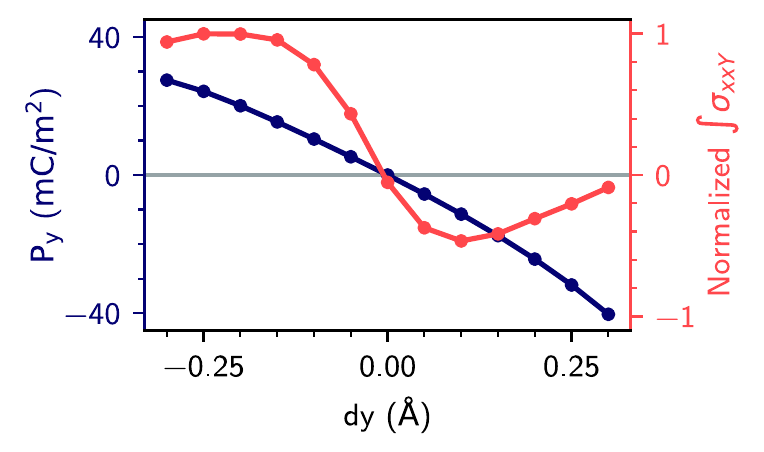}
\caption{Change in $y$-axis polarization and integrated shift current during \ch{Mo} displacement.}
\label{fig:pol}
\end{figure}

$2H$-\ch{MoS2} does not exhibit spontaneous polarization, but the optimal displacement identified above involves moving the different atomic species in opposite directions, causing polarization along the direction of displacement (Figure~\ref{fig:pol}). As is observed in group-IV monochalcogenides, the integrated shift current response is linearly related to polarization at small displacements, while at larger displacements, the relationship is sublinear and has maxima at both positive and negative displacements.

The optimized displacement only affects atomic positions within the unit cell and does not involve changes to the lattice parameters. It therefore cannot be realized through mechanical strain (which in general induces no net polarization). However due to the relationship between polarization and electric field, the desired distortion can be driven by an external field. This is analogous to the thermo-photovoltaic effect, which works when the instantaneous atomic displacements from phonons are altered with temperature variation, except the applied electric field drives only a single distortion. We approximate the influence of an electric field on the structure using the Born effective charges of the distorted structure and find the field required to produce such a distortion by equating the \emph{ab initio} force on the structure and the force due to the electric field. Solving the force balance equation $F = Z* \cdot E$ in the plane of the monolayer indicates that an in-plane electric field of \SI{6.9e10}{\V/\m} could be used to realize the displacement outlined above. This finding suggests that external electric field is an additional avenue for engineering the shift current response of materials, in complement to mechanical methods explored in previous work.

A natural extension of this work is to group-IV monochalcogenides, which are also promising optoelectronic materials; \ch{SnS} in particular has found application in photovoltaics and photodetectors \cite{Antunez2011,Deng2012,Krishnamurthi2020}. These materials are ``softer" and have piezoelectric coefficients an order of magnitude larger than transition metal dichalcogenides \cite{Fei2015}. In addition, a coupling between polar distortion and shift current response has been demonstrated in these materials \cite{Rangel2017sctmmc}, thus it is possible that the optical response can be engineered using much lower fields than are discussed in this work. The maximum photovoltaic response of \ch{GeSe} was shown to occur very close to the spontaneous polarization.
Further study of different compositions and of distortions not confined to the polar axis are needed to predict whether the bulk photovoltaic response is susceptible to engineering with external electric fields.

In conclusion, using a gradient descent materials design strategy, we show that the shift current response of monolayer $2H$-\ch{MoS2} can be greatly enhanced with an atomic displacement aligned on the $y$-direction. This displacement increases the integrated $\sigma_{xxY}$ response more than ten-fold and can be realized experimentally by applying an external electric field. This result not only poses applied electric field as a new method to engineer the shift current BPVE response of materials, but it also shows the potential benefits of materials design---perhaps with the right perturbations, such large enhancements can be coerced from many materials.


\begin{suppinfo}
Details of computational methods, momentum-resolved shift vector and transition intensity maps.
\end{suppinfo}

\begin{acknowledgement}
This work has been supported by the Department of Energy, Office of Science, Office of Basic Energy Sciences, under grant number DE-FG02-07ER46431. Computational support was provided by the National Energy Research Scientific Computing Center (NERSC) of the U.S. DOE.
\end{acknowledgement}

\bibliography{refs}

\end{document}


\maketitle
\newpage


\section{Calculation details}

The \textsc{Quantum Espresso} package was used to perform density functional theory (DFT) calculations of the electronic structure using a generalized gradient approximation functional \cite{Giannozzi2009,Giannozzi2017,Perdew1996}. Calculations used norm-conserving pseudopotentials designed using the OPUIM package \cite{Rappe1990,Ramer1999}. Self-consistent calculations of the electron density were performed on an \num{12x12x1} Monkhorst-Pack grid \cite{Monkhorst1976}. Wavefunctions were calculated non-self-consistently on a denser \num{96x96x1} grid. A vacuum gap of \SI{12}{\angstrom} between layers was included to reduce spurious inter-layer interactions. The shift current and related quantities were calculated according to the method described by Young and Rappe \cite{Young2012}. Born effective charge tensors were calculated using density functional perturbation theory (DFPT), and polarization was calculated using the Berry phase method \cite{King-Smith1993}. The convergence threshold was \SI{e-11}{Ry/cell} for DFT calculations and \SI{e-18}{Ry/cell} for DFPT calculations.


\newpage
\section{Momentum-resolved shift vector and transition intensity}

\begin{figure*}
    \centering
    \includegraphics{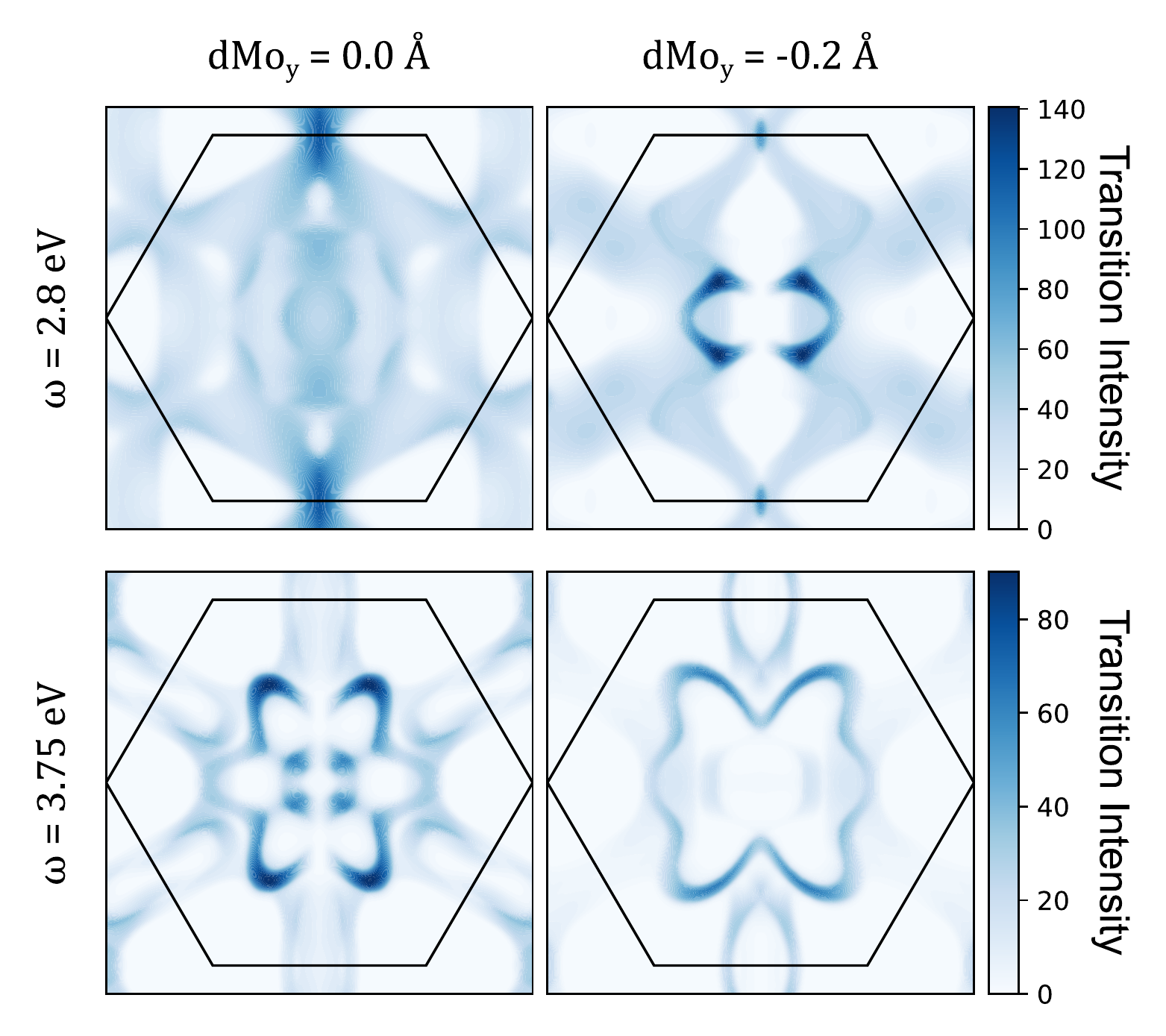}
    \caption{Momentum-resolved transition intensity of the equilibrium and distorted structures of \ch{MoS2} for transitions near $\omega =$ \SIlist{2.8; 3.75}{\eV}.}
    \label{fig:bz_tr}
\end{figure*}

\begin{figure*}
    \centering
    \includegraphics{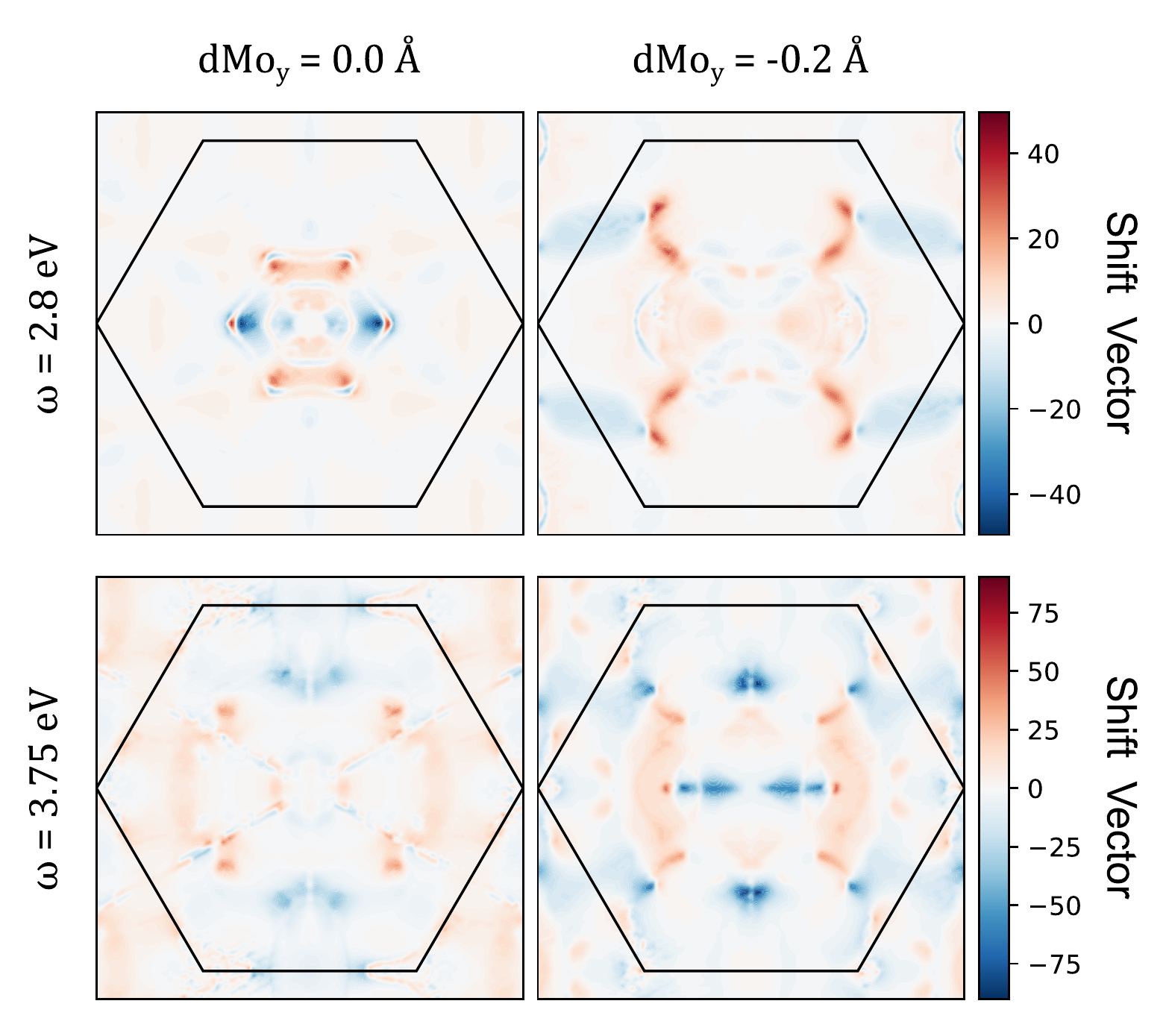}
    \caption{Momentum-resolved shift vector of the equilibrium and distorted structures of \ch{MoS2} for transitions near $\omega =$ \SIlist{2.8; 3.75}{\eV}.}
    \label{fig:bz_sv}
\end{figure*}


\clearpage

\bibliography{refs}